\documentclass[prl,amsmath,twocolumn,showpacs,superscriptaddress]{revtex4}
\usepackage{graphicx}
\usepackage{psfrag}
\usepackage[english]{babel} %indica la lingua usata
\usepackage{amsmath,amsthm,amssymb,amsfonts,latexsym} %libreria per simboli vari
\usepackage[latin1]{inputenc} %pkt per mettere gli accenti
\newlength{\picwidth}
\begin{document}
\newcommand{\beq}{\begin{equation}}
\newcommand{\eeq}{\end{equation}}
\newcommand{\ben}{\begin{eqnarray}}
\newcommand{\een}{\end{eqnarray}}
\newcommand{\bea}{\begin{array}}
\newcommand{\eea}{\end{array}}
\newcommand{\om}{(\omega )}
\newcommand{\bef}{\begin{figure}}
\newcommand{\eef}{\end{figure}}
\newcommand{\leg}[1]{\caption{\protect\rm{\protect\footnotesize{#1}}}}
\newcommand{\ew}[1]{\langle{#1}\rangle}
\newcommand{\be}[1]{\mid\!{#1}\!\mid}
\newcommand{\no}{\nonumber}
\newcommand{\etal}{{\em et~al }}
\newcommand{\geff}{g_{\mbox{\it{\scriptsize{eff}}}}}
\newcommand{\da}[1]{{#1}^\dagger}
\newcommand{\cf}{{\it cf.\/}\ }
\newcommand{\ie}{{\it i.e.\/}\ }
\title{Coherent transport in multi-branch quantum circuits}
\author{A.~Ziletti}
\affiliation{Dipartimento di Matematica e Fisica, Universit\`a Cattolica, via Musei 41, 25121 Brescia, Italy}
\author{F.~Borgonovi}
\affiliation{Dipartimento di Matematica e Fisica, Universit\`a Cattolica and Interdisciplinary Laboratories for Advanced Materials Physics, via Musei 41, 25121 Brescia, Italy}
\affiliation{I.N.F.N., Sezione di Pavia, Italy}
\author{G.~L.~Celardo}
\affiliation{Dipartimento di Matematica e Fisica, Universit\`a Cattolica and Interdisciplinary Laboratories for Advanced Materials Physics, via Musei 41, 25121
Brescia, Italy}
\affiliation{I.N.F.N., Sezione di Pavia, Italy}
\author{F.~M. Izrailev}
\affiliation{Instituto de F\'{\i}sica, Universidad Aut\'{o}noma de
Puebla, Apartado Postal J-48, Puebla, Pue. 72570, M\'{e}xico}
\affiliation{NSCL and Department of Physics and Astronomy, Michigan State
University, East Lansing, Michigan 48824-1321, USA}
\author{L. Kaplan}
\affiliation{Department of Physics, Tulane University, New Orleans, Louisiana 70118, USA}
\author{V.~G.~Zelevinsky}
\affiliation{NSCL and Department of Physics and Astronomy, Michigan State
University, East Lansing, Michigan 48824-1321, USA}

\begin{abstract}
An open multi-branch quantum circuit is considered from the viewpoint of coherent electron or wave transport,
both with and without intrinsic disorder. Starting with the closed system, we give analytical conditions for
the appearance of two isolated localized states out of the energy band. In the open system, using the method
of the effective non-Hermitian Hamiltonian, we study signal transmission through such a circuit, with
an important result of a long lifetime of localized states. When the average level width becomes comparable
to the mean level spacing, the super-radiant transition occurs. In the case of on-site disorder we find
an analytical estimate, confirmed by numerical data, for the robustness of the isolated states and their role
in transport processes.
\end{abstract}
\date{\today}
\maketitle

%%%%%%%%%%%%%%%%%%%%%%%%%%%%%%%%%%%%%%%%%%%%%%%%%%%%%%%%%%%%%%%%%%%%%%%%%%%%%%%%%%%%
\section{Introduction}

The development of quantum informatics requires better understanding of the general problem of quantum signal
transmission through discrete structures of interacting quantum elements, such as quantum dots
\cite{hanson07,kitavittaya09,granger10}, or molecular \cite{solomon08,rebentrost09} and Josephson junctions
\cite{fulton89}. There is growing theoretical and experimental interest in arrangements more complicated than
a simple one-dimensional chain, including $Y$- and $T$-shaped structures \cite{santos03,giuliano09}, tetrahedral
qubits \cite{feigelman04}, connected benzene rings \cite{stafford07}, crossed chains \cite{yan11}, two- and
three-dimensional lattices \cite{mancini07,PRB10}, as well as graphs with the violation of time-reversal
invariance at the junctions, supposedly with the aid of magnetic fields \cite{bellazzini09}. In all cases,
one has to deal with a quantum system with intrinsic stationary states that become unstable when the system
``opens" to the external world as a part of a transmission network.

The method of the effective non-Hermitian Hamiltonian, borrowed from nuclear physics \cite{zele89}, is a powerful
and efficient tool for theoretical analysis of open quantum systems, as was shown by applications to one-dimensional
structures \cite{SZAP92,rotter,VZ05,sorathia09,celardo09,PRB10}, covering both regular and chaotic internal dynamics;
see the recent review of the method and its various adaptations \cite{auerbach11}. In this work we consider
the $M$-branch circuit in the form of $M>2$ one-dimensional tight-binding chains with a common vertex at
the central point. This system is a simple discrete example of quantum mechanics in a non-trivial space with
self-crossings, singular points or surfaces where quasi-bound states (evanescent waves) may emerge even for
open boundary conditions. Our interest is not only in the structure of the energy spectrum and eigenfunctions for
closed samples, which was earlier discussed for different purposes, see for example, the applications to
Bose-Einstein condensation and temperature effects in optical networks \cite{brunelli04}, but mainly in
the transport characteristics for the case when the chains are connected to an environment.

\vspace{0.3cm}
\begin{figure}[!ht]
\vspace{0cm}
\includegraphics[width=6cm]{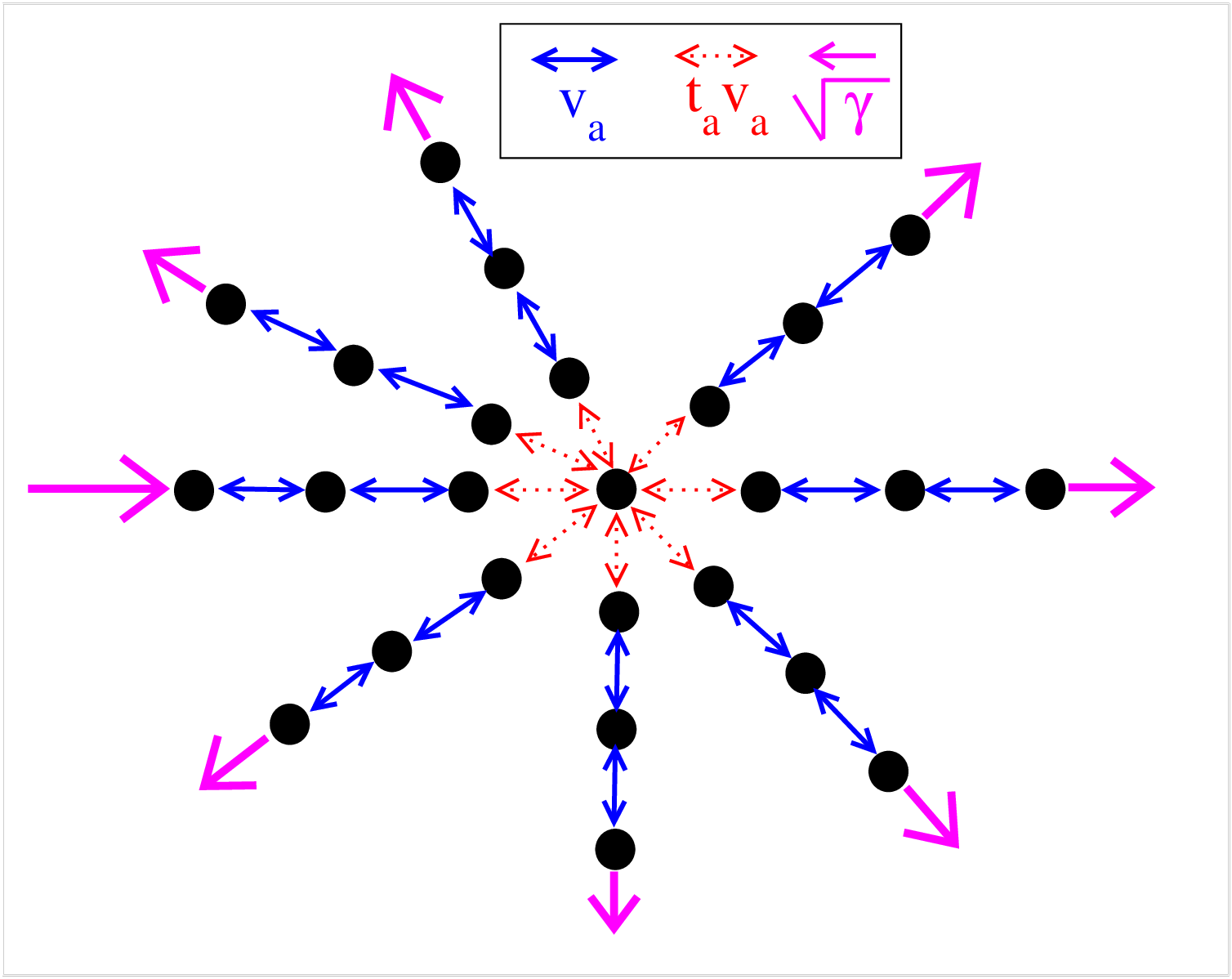}
\caption{{\footnotesize (Color online) The $M$-branch circuit with the coupling at the origin.
}}
\label{uno}
\end{figure}

%%%%%%%%%%%%%%%%%%%%%%%%%%%%%%%%%%%%%%%%%%%%%%%%%%%%%%%%%%%%%%%%%%%%%%%%%%%%%%%%%%%%
\section{Model}

We start with the discussion of the closed system and its energy spectrum. In our model (see Fig.~\ref{uno}),
each of the $M$ branches consists of $N_a$ sites ($a=1,...,M$) along which a particle/wave can propagate through
the structure. The hopping amplitude $v_a$ between nearest neighbor sites is constant within each branch, while
the coupling to the origin is given by $t_a v_a$. Thus, the Hamiltonian of the closed system is
\begin{align}
\label{eq:hamil}
\begin{split}
\mathsf{H} =& \sum_{a=1}^M \sum_{n=1}^{N_a-1} v_a
\Bigl(\vert a, n\rangle \langle a, n +1\vert +
 \vert a, n+1 \rangle \langle a, n \vert\Bigr) \\
 &+ \sum_{a=1}^M t_a v_a  \Bigl( \vert 0 \rangle \langle a,
1 \vert + \vert a, 1 \rangle \langle 0 \vert \Bigr)\,.
\end{split}
\end{align}

In what follows, unless stated otherwize, we set  $v_a = v$ and $N_a = N$, so that there are $K=MN + 1$ sites
in total, including the vertex. The Schr\"{o}dinger equation for a stationary state with energy $E$ reduces to
a set of algebraic equations for the site amplitudes $C_{n}^{a}$, where $a=1, \ldots, M$, and $n$ labels the sites
in each chain,
\begin{eqnarray}\label{eq:system_rip}
\sum_{a=1}^M t_a v C_1^a &=& E C_0\,; \,\,\,\,\,\,\,\,\,\, v C_2^a + t_a v  C_0  = E C_1^a \,;
\nonumber\\[6pt]
v (C_{n-1}^a+C_{n+1}^a) &=& E C_n^a \;\;\;\; {\rm for} \;\;\;\; 2 \le n \le N\,,
%v C_{N-1}^a = E C_N^a .
\end{eqnarray}
with the boundary conditions  $C_{N+1}^a = 0$.

%%%%%%%%%%%%%%%%%%%%%%%%%%%%%%%%%%%%%%%%%%%%%%%%%%%%%%%%%%%%%%%%%%%%%%%%%%%%%%%%%%%%
\section{Spectrum and eigenfuctions}

For $C_0 \ne 0$, the $a$-dependence of the amplitudes is given by $C_{n}^{a}=vt_{a}X_{n}$ with  $n=1, \ldots, N$, while $C_0$ and the $a$-independent amplitudes $X_{n}$ obey a tri-diagonal homogeneous set of $N+1$ linear equations.
The secular equation for $E\ne \pm 2v$ has the form,
\begin{equation}
(E^{2}+E\epsilon_{+}-v^{2}Q)\epsilon_{-}^{N}-(E^{2}+
E\epsilon_{-}-v^{2}Q)\epsilon_{+}^{N}=0\,,         \label{eq:polyn_ttot}
\end{equation}
with the control parameter $Q$\,,
\begin{equation}
\label{key}
Q =\sum_{a=1}^M t_a^2\,,
\end{equation}
and $\epsilon_\pm(E) = (-E \pm \sqrt{E^2-4v^2})/2$.  Note that for even $N$ the value $E=0$ is also a solution of
Eq.~(\ref{eq:polyn_ttot}). An analysis of Eq.~(\ref{eq:polyn_ttot}) shows that the energy spectrum consists of a set
of $MN-1$ eigenvalues within the energy band $|E|<2v$ (with eigenfunctions extended over the $M$ branches of
the circuit), and two additional eigenvalues with $|E|>2v$, for sufficiently large $M$ or sufficiently large
couplings $t_a$.

We first consider the two states with the eigenvalues outside of the energy band (sometimes such states are called
``hidden"). These states turn out to be strongly localized at the origin of the circuit provided the parameter $Q$
is large enough. In the limit $N \gg 1$, the energies of these states can be found from Eq.~(\ref{eq:polyn_ttot})
by considering its largest and smallest roots. For $Q > 2 $ one obtains
\begin{align}
|\mathcal{E}_{\rm loc}| =\frac{Q}{\sqrt{Q-1}}\,v \,,
\end{align}
which generalizes the results found in Refs.~\cite{pastawski01}.

In order to find the structure of the corresponding eigenfunctions from
Eq.~(\ref{eq:system_rip}),  we obtain, after lengthy calculations,
the following relation, valid for any finite $N$:
\begin{equation}
\label{eq:ratio_gen2}
\left \vert \dfrac{C^a_{n+1}}{C^a_n} \right \vert^2 =
\frac{1}{Q-1}\, \mathcal{B}_{n}(Q,N), \quad n=1,\ldots,N-1\,,
\end{equation}
where we have defined the boundary factor,
\begin{equation}
\label{eq:boundary}
\mathcal{B}_{n}\left (Q,N\right ) = \left \vert \dfrac{1 - \left (Q-1\right )^{-\left ( N-n \right )}}{1-
\left (Q-1\right )^{-\left ( N+1-n \right )}} \right \vert^2\,,
\end{equation}
which differs from 1 only for $n \approx N$.
Neglecting for the time being this boundary effect
(which however will be crucial for an open model, see below), we obtain
%from Eq.~(\ref{eq:ratio_gen2}),
\begin{align}
\label{eq:decay_t_4}
\left \vert \dfrac{C_n^a}{C_0} \right \vert^2 =
\dfrac{t_a^2}{Q-1} e^{-\left (n-1 \right )/\xi}, &&
n=1,\ldots, N; && Q >  2\,.
\end{align}
Here $\xi$ is the localization length,
\begin{equation}
\label{eq:loclength}
\xi = \frac{1}{\ln(Q-1)} \,,
\end{equation}
of the two states outside the energy band, with the peak located at the origin of the circuit. Note that these states
are spread over {\it all branches} in proportion to the couplings $t_a^2$.

For $C_0=0$, one has standard extended Bloch states in each branch. However, the $M$ solutions are linked through
the first equation of (\ref{eq:system_rip}), yielding $M-1$ independent degenerate states for each $C_0=0$ eigenvalue.
As a result, for the symmetric case of equal branch lengths $N_a=N$, we have two isolated localized states, $N$ sets
of $M-1$ extended degenerate
%$(M-1)N$ extended
Bloch states with $C_0=0$, and $(N-1)$ extended non-degenerate Bloch states with $C_0 \ne 0$, altogether $K=NM+1$ states.

%%%%%%%%%%%%%%%%%%%%%%%%%%%%%%%%%%%%%%%%%%%%%%%%%%%%%%%%%%%%%%%%%%%%%%%%%%%%%%%%%%%%
\section{Coupling to continuum}

Coming to the main goal of our study, now we consider the same circuit coupled to the environment by attaching
the last site of each branch to an external channel, similarly to what has been done in
Refs.~\cite{SZAP92,VZ05,celardo09,PRB10}. The open system is described by an effective non-Hermitian Hamiltonian
\cite{zele89,auerbach11},
\begin{align}
\label{eq:h_open_s}
\begin{split}
\mathcal{H} = \mathsf{H} -  \dfrac{i}{2}\gamma
\mathsf{W} ; \,\,\,\,\,\, \,\,\,
\mathsf{W} =\sum_{a=1}^M A_N^{a} |a,N\rangle \langle a,N| A_N^{a\ast} \,.
\end{split}
\end{align}
Here $\mathsf{H}$ is given by Eq.~(\ref{eq:hamil}), while the coupling to the continuum via the ends of branches is
characterized by the parameter $\gamma$ and the matrix $\mathsf{W}$. The matrix $\mathsf{W}$ is constructed out
of the transition amplitudes $A_N^{a}$ between intrinsic states $|a,N \rangle$ and continuum states $|a, E\rangle $.
We set $A_N^{a}=1$, so that the strength of the coupling is controlled entirely by the parameter $\gamma$.

The lifetime $\tau_{\rm \, loc}$ of the two localized states can be estimated via the imaginary part
$-\Gamma_{{\rm loc}}/2$ of the corresponding eigenvalues $\mathcal{E}_{{\rm loc}}$ of the complex Hamiltonian,
$\tau_{\rm \, loc} = \Gamma_{{\rm loc}}^{-1}$. For small $\gamma$, the resonance width is determined by the spatial
overlap of the localized states with the edges,
\begin{equation}
\Gamma_{{\rm loc}} =\gamma \sum_{a=1}^M \vert \langle a,N_a \vert \psi_{{\rm loc}}\rangle \vert^2,
\end{equation}
see Refs.~\cite{zele89,SZAP92,rotter}. Taking into account Eqs.~(\ref{eq:ratio_gen2}, \ref{eq:boundary},
\ref{eq:loclength}) with $\vert C_{0} \vert^2$ found from the normalization, for $N, Q \gg 1$ we have
\begin{equation}
\label{eq:tau_loc55}
\tau_{\rm \, loc} =  2 \dfrac{\left (Q-1\right )^{N+1}}{\gamma \, Q(Q-2)} \simeq \frac{2}{\gamma}
\exp \left(\frac{N-1}{\xi}\right).
\end{equation}

Considering now the extended states inside the energy band, two different regimes can be distinguished
\cite{zele89,SZAP92} as a function of $\gamma$. At weak coupling, all these states are similarly affected
by the continuum coupling and acquire widths proportional to $\gamma$. For large $\gamma$, only $M$
``{\sl super-radiant}" states have a width proportional to $\gamma$, while the widths of the remaining
(``{\sl trapped}") states fall off as $1/\gamma$. In order to find the critical value of the parameter
$\gamma$ corresponding to the super-radiant transition, we analyze the average value $\langle \Gamma\rangle$
of the $(MN+1)-M$ narrow widths  as a function of the rescaled coupling $\gamma/v$. At a critical value
$\gamma_{{\rm cr}}$, the average width $\langle \Gamma\rangle$ peaks and begins to decrease.

\vspace{0.3cm}
\begin{figure}[!ht]
\vspace{0cm}
\includegraphics[width=6cm]{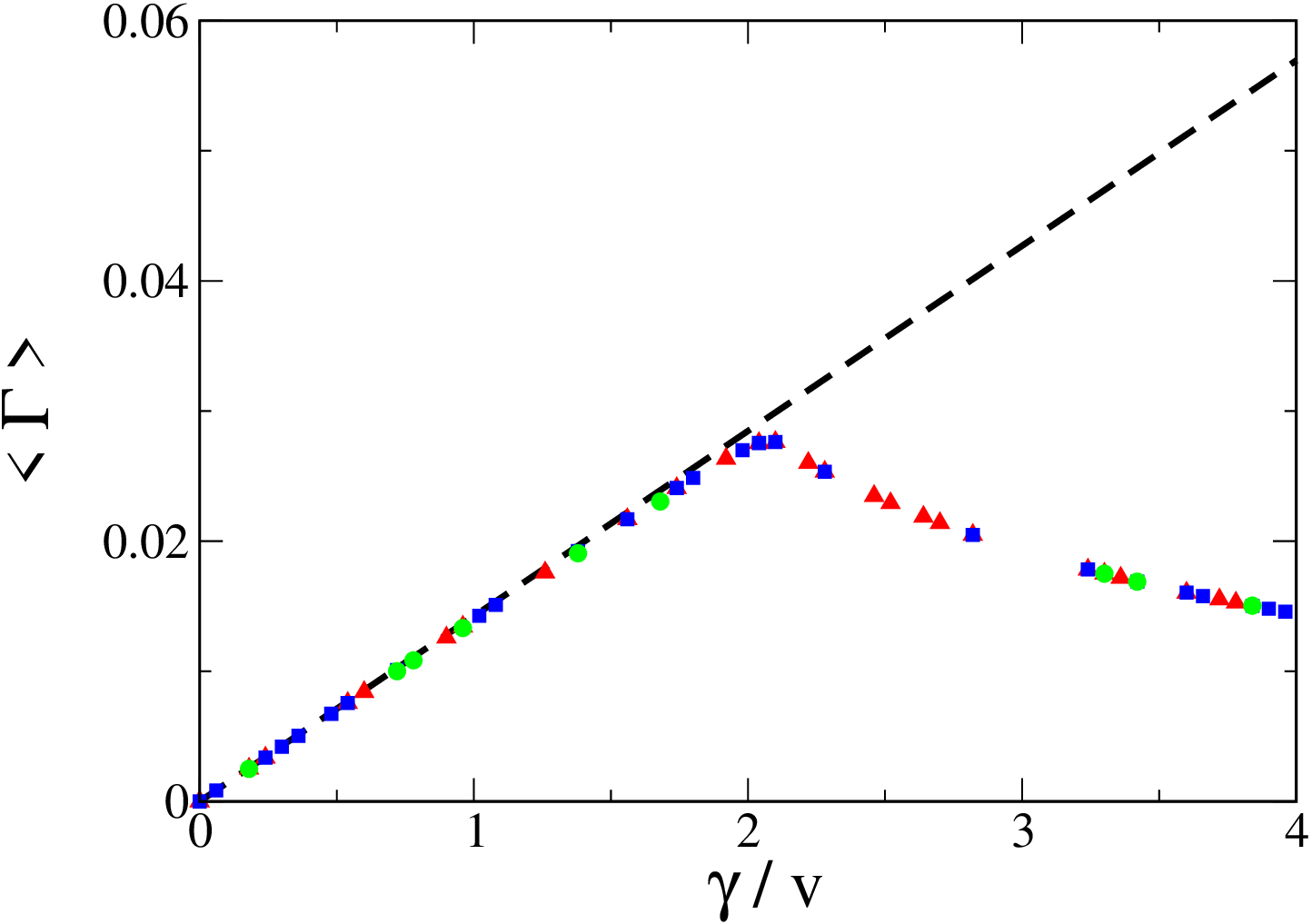}
\caption{{\footnotesize (Color online) The dimensionless average width $\langle \Gamma \rangle$ as
a function of $\gamma/v$. We consider a)  $M=4$ and $N=70$ with $t_a = 1$ for all branches (red triangles);
b) $M=4$ and $N=70$ with different couplings $t_{1}=200$, $t_{2}=100$,
$t_{3}=20$, and $t_{4}=50$ (blue squares);
and c) $M=4$ with different numbers of sites,
$N_1=55, N_2=80, N_3=58, N_4=87$, and all $t_a=1$ (green circles).
The dashed line corresponds to the average over all widths,
while the symbols are obtained by averaging over the $K-4$ smallest widths,
where $K=281$ is the total number of sites in all cases.}}
\label{fig:graf_crit_g}
\end{figure}

One can evaluate  $\gamma_{{\rm cr}} $ using the following
criterion \cite{SZAP92,celardo09}: the transition occurs
when $\langle \Gamma\rangle$ becomes of the order of the mean level
spacing $D$ of the Hamiltonian for the closed system.
This is particularly easy
for the special case of equal coupling $t_a=t$
and equal number of sites in each branch $N_a=N$; in this case
it  is convenient to define an effective mean level spacing as
$D_{{\rm eff}} \approx 4 v/ (2N+1) \simeq 2v/N$,
which takes into account the level degeneracy.
On the other hand, the average width is given by
$
\langle \Gamma\rangle \approx M\gamma/(MN+1) \simeq \gamma/N
$,
so that  $\gamma_{{\rm cr}}=2v$ independently of $M$, $N$, and $t_a$.
This result is numerically confirmed in  Fig.~\ref{fig:graf_crit_g},
where it is shown to be valid even in the more general case of different
coupling $t_a$ and different number of sites $N_a$.

%%%%%%%%%%%%%%%%%%%%%%%%%%%%%%%%%%%%%%%%%%%%%%%%%%%%%%%%%%%%%%%%%%%%%%%%%%%%%%%%%%%%
\section{Transmission}

\begin{figure}[!ht]
\vspace{0cm}
\includegraphics[width=6cm]{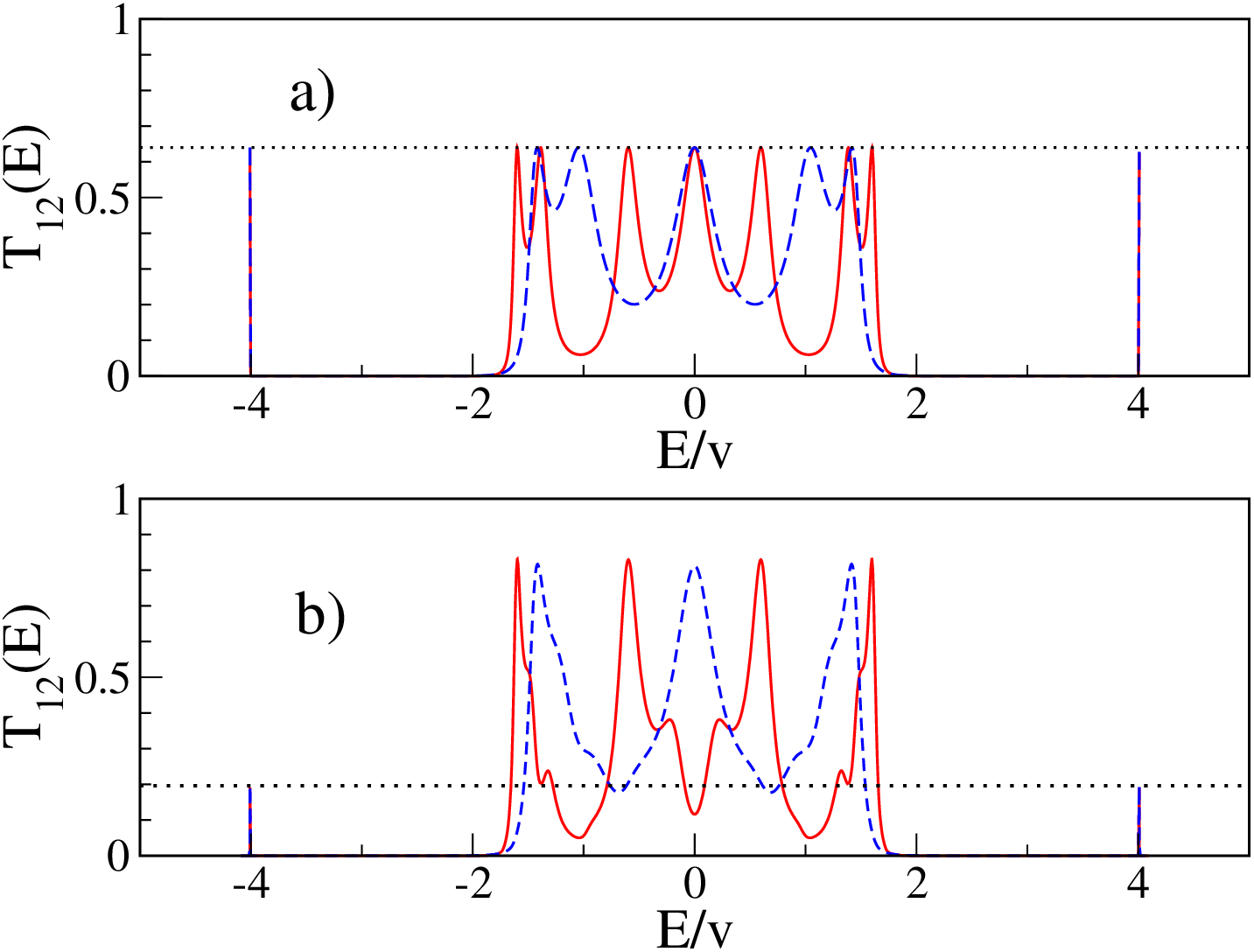}
\caption{{\footnotesize (Color online)
Transmission coefficients $T_{12}$ as a function of energy,
for $M=4$,  $t_1 = 3$, $t_2=2$, and $t_3=t_4=1 $.
Full  red curves stand for $\gamma=0.6$, while dashed
blue curves correspond to $\gamma=5$.
a) symmetric case $N=4$; b) asymmetric case $N_1=4, N_2=4, N_3=3,
N_4=5$.
Horizontal dotted lines are the theoretical results given by
Eq.~(\ref{eq:trasm_ab}) and its generalization to unequal branch lengths.
}}
\label{trass}
\end{figure}

Maximum transmission occurs at the super-radiant transition, as
happens in one-dimensional chains \cite{celardo09}.
Moreover the above analysis allows one to understand
generic properties of the transmission between different branches.
The data in Fig.~\ref{trass} demonstrate the energy dependence
of the transmission coefficient $T_{ab}$ between  two channels
 \textit{b} and \textit{a}. One can see that with an increase
 of $\gamma$ the resonances corresponding to the energies from
the bulk of spectrum begin to overlap, in contrast to the two very
 narrow resonances located out of the energy band.
These quasi-bound resonances remain extremely stable even for a very
strong coupling to continuum.

\vspace{0.3cm}

Of special interest is the value of the transmission coefficient
$T_{ab}$ for the resonances outside the band. Since these resonances
are very narrow, one can estimate $T_{ab}(\mathcal{E}_{\rm  loc})$
\cite{SZAP92,angelotesi} in terms of the edge components of the localized
state $\psi_{{\rm loc}}$ in channels $a$ and $b$,
\begin{equation}
T_{ab}(\mathcal{E}_{{\rm loc}}) = \left \vert \dfrac{\langle a,
N_a \vert \psi_{{\rm loc}} \rangle \sqrt{\gamma}
\langle \psi_{{\rm loc}} \vert b,
N_b \rangle \sqrt{\gamma} }{(i/2)\Gamma_{{\rm loc}}} \right \vert^2\,.
\end{equation}
which can be computed
from Eqs (\ref{eq:ratio_gen2}, \ref{eq:boundary}):
\begin{equation}
\label{eq:full}
|\langle a,N_a \vert \psi_{{\rm loc}}\rangle|^2=|C_0|^2t_a^2 e^{- N_a/ \xi}
\prod_{n=1}^{N_a-1} \mathcal{B}_{n}\left (Q,N_a\right )\,.
\end{equation}
In  the special case of equal-length branches this result assumes
the particularly simple Hauser-Feshbach form:
\begin{equation}
\label{eq:trasm_ab}
T_{ab}(\mathcal{E}_{{\rm loc}}) = \dfrac{4 t_a^2 t_b^2}{Q^2}, \quad a \neq b.
\end{equation}
This  relation is in good agreement with
the data of Fig.~\ref{trass}a, as indicated by the dotted horizontal
line. The value of $T_{ab}(\mathcal{E}_{{\rm loc}})$
for narrow resonances depends on the hopping elements $t_a$ only,
and is independent of the coupling
strength $\gamma$ for equal values of the
couplings, $\gamma_a=\gamma$. The numerical data also
indicate that, for  equal-length branches,
the maximal value of the transmission coefficient for $|E|<2$ is given by the
same expression (\ref{eq:trasm_ab}), while this does not happen
when branches have different length,
see Fig.~\ref{trass}~b). As $\gamma\rightarrow 0$, the lifetime
of the localized states becomes very long.

%%%%%%%%%%%%%%%%%%%%%%%%%%%%%%%%%%%%%%%%%%%%%%%%%%%%%%%%%%%%%%%%%%%%%%%%%%%%%%%%%%%%
\section{Introducing disorder}

A key consideration in important practical applications is the influence of disorder.
For this reason we study the circuit with diagonal disorder by adding the term $V$,
\begin{equation}
V = \sum_a \sum _n  \epsilon_{a,n}  | a, n  \rangle \langle a, n |,
\end{equation}
to the non-Hermitian Hamiltonian (\ref{eq:h_open_s}). Here the site
energies $\epsilon_{a,n}$ are random numbers uniformly distributed
in the interval $[- W/2, W/2]$. According to the theory of disordered systems,
for $N \rightarrow \infty$ all states within the energy band become
exponentially localized with a localization length $\propto (v/W)^2$.
For the isolated localized states, a weaker dependence on disorder
can be expected, and it is more convenient to define a localization
length $\ell_{{\rm loc}}$ through the inverse participation number,
\begin{equation}
\ell_{{\rm loc}} = \left (
|\langle 0 | \psi_{{\rm loc}} \rangle|^4 +
\sum_{a=1}^{M}  \sum_{n=1}^N \vert \langle a, n | \psi_{{\rm loc}} \rangle\vert^4
\right) ^{-1}.
\end{equation}
Without disorder and for large $N$, the value of $\ell_{{\rm loc}}$ for the isolated
states can be estimated as
\begin{equation}
\ell_{{\rm loc}} \approx \dfrac{4 Q (Q-1)^2}{(Q-2)\left
[ Q(Q-2)+\sum_{a=1}^M t_a^4  \right ]}.
\end{equation}

\begin{figure}[!ht]
\vspace{0cm}
\includegraphics[width=6cm]{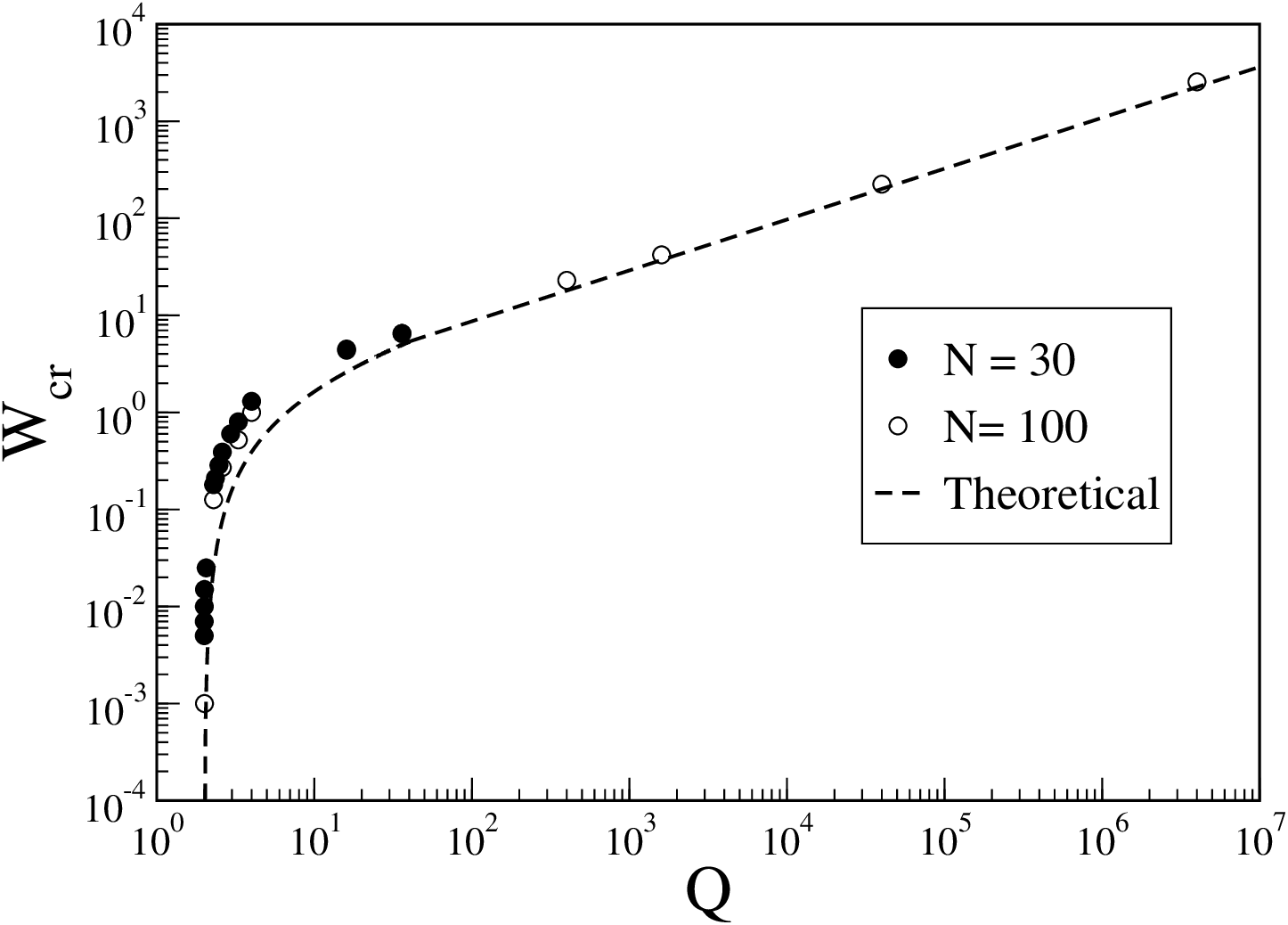}
\caption{{\footnotesize  Critical value $W_{{\rm cr}}$
in units where $v=1$, as a function of $Q= \sum_a t_a^2$ for $M=4$.
We choose for simplicity  $t_a= t $ for $a=1, \ldots, M$ , so that $Q=M t^2$.
Circles represent numerical data (full $N=30$, open $N=100$),
while the dashed line shows
the prediction (\ref{vuc}).}}
\label{WW}
\end{figure}

Numerical data confirm that the isolated states are  not affected
by disorder up to a critical disorder strength,  $W_{{\rm cr}}$,
that can be very large. Above $W_{{\rm cr}}$,
the localization length of the isolated states also begins
to decrease. This critical value can be estimated by assuming that it
corresponds to the intersection between the gap of size $\Delta$ emerging due
to disorder around the localized states and the bulk of the spectrum of width $2(2v+W/2)$.
One can estimate the value of $\Delta$ from the relation,
\begin{equation}
\Delta^2 = \overline{\langle \psi_{{\rm loc}} \vert V \vert \psi_{{\rm loc}} \rangle^2} =
\dfrac{W^2}{12}\,\ell_{{\rm loc}}^{-1}\,,
\end{equation}
where the average is taken over sites $n$ and over the disorder. The critical value
$W_{{\rm cr}}$ for large $N$ is thus obtained by
equating the half-width of the density of states ($2v + W/2$) to the minimal
possible energy of the isolated states due to random fluctuations:
\begin{equation}
\dfrac{Q v }{\sqrt{Q -1}} - \Delta  \approx  2v+ \dfrac{W}{2}\,,
\end{equation}
which yields
\begin{equation}
\begin{array}{lll}
W_{{\rm cr}} &= 2 v \left ( \dfrac{Q}{\sqrt{Q-1}} - 2 \right )
\times \\
&\times \left ( 1 + \sqrt{\dfrac{\left (Q-2\right )\left [
Q \left (Q-2 \right ) +
\sum_{a=1}^M t_a^4 \right ]}{12 Q
\left (Q-1 \right )^2}} \right )^{-1}.
\end{array}
\label{vuc}
\end{equation}
Despite the cumbersome appearance of Eq.~(\ref{vuc}), it admits two interesting
limits: assuming for simplicity $t_{a}=t$ for all $a$, it is easy to see that
for $Q \approx 2 $, $W_{{\rm cr}}/v \simeq (Q-2)^2/2$, while for $Q \to \infty$,
one has $W_{{\rm cr}}/v \simeq k \sqrt{Q}$, where the constant
$k = 2/(1+\sqrt{(M+1)/(12 M)})$ depends only on the number of branches, and
$k\approx 1.55$ for large $M$.
Note that it is impossible to have delocalized isolated states
along with localized states inside the band for any disorder $W$.
The expression (\ref{vuc}) is compared with
numerical data for $W_{{\rm cr}}$ defined as the point where the localization length
$\ell_{{\rm loc}}$ of the isolated localized state begins to decrease with an increase
of disorder. The data in Fig.~\ref{WW} show quite good agreement with this estimate
over many orders of magnitude with no fitting parameters.

\vspace{0.3cm}

\section{Summary}

As an example of a non-trivial quantum network, we have studied the properties
of an open circuit with $M$ branches coupled to each other through one common point.
The transmission properties of the open model were studied in relation to
the structure of the energy spectrum and eigenfunctions of the parent closed system.
The method of the effective non-Hermitian Hamiltonian allows one to derive the exact
solution of the problem. Our main interest was in the energies and localization lengths
of special eigenstates located outside the crystal energy band and strongly localized at
the junction. It was shown that, in the presence of coupling to the continuum, these
states typically become narrow resonances with a very large lifetime. This fact may be
important for the fabrication of new kinds of electron nanostructures, waveguides,
antennas, and lasing devices with a large quality factor. Another possible application
of the $M$-branch circuits follows from the expression (\ref{eq:trasm_ab}),
which points out the possibility of controlling and distributing energy
incoming via one branch into all others. We have also shown the negligible
influence of disorder in the branches on these special eigenstates. In our consideration
the continuum coupling $\gamma$ was taken as a constant parameter. In real
arrangements, the circuit can be connected to transmitters or particle reservoirs.
Then $\gamma$ can depend on the signal energy and density of states in the reservoirs.
Along with that, the continuum coupling will acquire a real part (dispersive integral)
that should be added to the intrinsic Hamiltonian in Eq. (10). We hope to consider this
more complicated situation elsewhere.

\vspace{0.5cm}
{\small G.L.C is grateful to H. Pastawski for useful discussions.
This work has been supported by Regione Lombardia and CILEA Consortium through a LISA Initiative (Laboratory for Interdisciplinary Advanced Simulation) 
2011 grant [link: http://lisa.cilea.it ].
This work was supported in part by the NSF under Grants No. PHY-0758099,
PHY-1068217, and PHY-0545390. F.M.I. is thankful to VIEP of the BUAP (Puebla,
Mexico) for financial support.
Support by the grant D.2.2 2010
(Calcolo ad alte prestazioni)  from Universit\`a Cattolica
 is also acknowledged.}

%\end{multicols}

\begin{thebibliography}{99}

\bibitem{hanson07} R. Hanson, L. P. Kouwenhoven, J. R. Petta, S. Tarucha, and
L. M. K. Vandersypen, Rev. Mod. Phys. {\bf 79}, 1217 (2007).

\bibitem{kitavittaya09} S. Kitavittaya, A. Rastelli, and O. G. Schmidt, Rep. Prog.
Phys. {\bf 72}, 046502 (2009).

\bibitem{granger10} G. Granger {\sl et al.}, Phys. Rev. B {\bf 82}, 075304 (2010).

\bibitem{rebentrost09} P. Rebentrost, M. Mohseni, I. Kassal, S. Lloyd, and A. Aspuru-Guzik,
New J. Phys. {\bf 11}, 033003 (2009).

\bibitem{solomon08} G. C. Solomon {\sl et al.}, J. Chem. Phys. {\bf 129}, 054701 (2008).

\bibitem{fulton89} T. A. Fulton {\sl et al.}, Phys. Rev. Lett. {\bf 63}, 1307 (1989).

\bibitem{santos03} L. F. Santos and M. I. Dykman, Phys. Rev. B {\bf 68}, 214410 (2003).

\bibitem{giuliano09} D. Giuliano and P. Sodano, Nucl. Phys. {\bf B811}, 395 (2009).

\bibitem{feigelman04} M. V. Feigel'man, L. B. Ioffe, V. B. Geshkenbein, P. Dayal, and
G. Blatter, Phys. Rev. B {\bf 70}, 224524 (2004).

\bibitem{stafford07} C. A. Stafford, D. M. Cardamone, and S. Mazumdar, Nanotechnology
{\bf 18}, 424014 (2007).

\bibitem{yan11} H. Yan {\sl et al.}, Nature {\bf 470}, 240 (2011).

\bibitem{mancini07} F.P. Mancini, P. Sodano, and A. Trombettoni, Int. J. Mod. Phys.
B {\bf 21}, 1923 (2007).

\bibitem{PRB10} G. L. Celardo, A. M. Smith, S. Sorathia, V. G. Zelevinsky,
R. A. Sen'kov, and L. Kaplan, Phys. Rev. B {\bf 82}, 165437 (2010).

\bibitem{bellazzini09} B. Bellazzini, M. Mintchev, and P. Sorba, Phys. Rev. B {\bf 80},
245441 (2009).

\bibitem{zele89} V. V. Sokolov and V. G. Zelevinsky, Nucl. Phys. {\bf A504},
562 (1989).

\bibitem{SZAP92} V. V. Sokolov and V. G. Zelevinsky, Ann. Phys.
(N.Y.) {\bf 216}, 323 (1992).

\bibitem{rotter} A. F. Sadreev and I. Rotter, J. Phys. A {\bf 36}, 11413 (2003).

\bibitem{VZ05} A. Volya and V. Zelevinsky, AIP Conference Proceedings
{\bf 777}, 229 (2005).

\bibitem{sorathia09} S. Sorathia, F. M. Izrailev, G. L. Celardo, V. G. Zelevinsky,
and G. P. Berman, EPL {\bf 88}, 27003 (2009).

\bibitem{celardo09} G. L. Celardo and L. Kaplan, Phys. Rev. B {\bf 79},
155108 (2010).

\bibitem{auerbach11} N. Auerbach and V. Zelevinsky, Rep. Prog. Phys.
{\bf 74}, 106301 (2011).

\bibitem{brunelli04} I. Brunelli, G. Giusiano, F.P. Mancini, P. Sodano, and
A. Trombettoni, J. Phys. B {\bf 37}, S275 (2004).

\bibitem{pastawski01} J. L. D'Amato, H. M. Pastawski, and J. F. Weisz,
Phys. Rev. B {\bf 39}, 3554 (1989);
H. M. Pastawski and E. Medina, Rev. Mex. Fis. {\bf 47}, 1 (2001).

\bibitem{angelotesi} A. Ziletti, Master's Thesis,
Universit\`a Cattolica, Brescia, Italy (2011), 	arXiv:1109.0727v1 [cond-mat.stat-mech].

\end{thebibliography}
\end{document}